    \documentstyle[12pt]{article}
\begin{document}
\baselineskip=14pt plus 1pt minus 1pt
\begin{center}
{\Large \bf  ``Beat'' patterns for the odd--even staggering in
octupole bands from a quadrupole--octupole Hamiltonian }
\end{center}
\medskip

\begin{center}

{\large Nikolay Minkov$^*$\footnote[1]{e-mail:
nminkov@inrne.bas.bg},
S. B. Drenska$^*$\footnote[2]{e-mail:
sdren@inrne.bas.bg},
P. P.Raychev$^{*}$\footnote[3]{e-mail:
raychev@inrne.bas.bg},
R. P. Roussev$^*$\footnote[4]{Deceased},
 and
Dennis Bonatsos$^\dagger$\footnote[5]{e-mail:
bonat@mail.demokritos.gr}\\
\medskip

$^*$  Institute for Nuclear Research and Nuclear Energy, \\
72 Tzarigrad Road, 1784 Sofia, Bulgaria\\
\medskip

$^\dagger$ Institute of Nuclear Physics, N.C.S.R. ``Demokritos'',\\
GR-15310 Aghia Paraskevi, Attiki, Greece}

\end{center}

\bigskip\bigskip

\begin{abstract}

We propose a collective Hamiltonian which incorporates the standard
quadrupole terms, octupole terms classified according to the
irreducible representations of the octahedron group,  a
quadrupole--octupole interaction, as well as a term for the
bandhead energy linear in $K$ (the projection of angular momentum
on the body-fixed $z$-axis). The energy is subsequently minimized
with respect to $K$ for each given value of the angular momentum
$I$, resulting in $K$ values increasing with $I$ within each band,
even in the case in which $K$ is restricted to a set of microscopically 
plausible values. 
We demonstrate that this Hamiltonian is able to reproduce a variety
of ``beat'' patterns observed recently for the odd--even staggering
in octupole bands of light actinides.

\end{abstract}

\bigskip


\bigskip

\newpage
\section{Introduction}

The properties of nuclear systems with octupole deformations
\cite{BM75} are of current interest due to increasing evidence for
the presence of octupole instabilities in various regions of the
nuclear table \cite{Ro88,Ahmad,BN96}. Furthermore, some ``beat''
patterns have been observed recently for the odd--even staggering
(the relative displacement of the odd levels with respect to the
positions at which they should have been located according to a fit
of the even levels by the formula $E(I)=A I(I+1)$, where $I$
denotes the angular momentum) in octupole bands of light actinides
\cite{DBoct00} based on recent experimental data
\cite{Cocks97,Cocks99}, calling for a study of the interactions
which could give rise to such shapes.

Various parametrizations of the octupole degrees of freedom
\cite{Ro90,Ham91,HBXZ91,WD99} already exist, being a useful tool
for understanding the role of the reflection asymmetry correlations
and for analyzing the collective properties of such systems. Some
important questions in this direction are: which are the collective
nuclear interactions that correspond to the different octupole
shapes and how do they determine the structure of the respective
energy spectra? Physically meaningful answers should be obtained by
taking into account the simultaneous presence of other collective
degrees of freedom, such as the quadrupole ones \cite{Ro82}.

In the present work we address the above problems by examining the
interactions that generate collective rotations in a system with
a simultaneous presence of octupole and quadrupole deformations.
The basic assumption of our consideration is that the rotational
motion of such a system can be interpreted in first
approximation as the motion of a body with a stable
quadrupole--octupole shape. In this respect our purpose is  to
examine how the nuclear system behaves under collective
rotations if the presence of stable quadrupole--octupole
deformations is assumed.

Based on the octahedron point symmetry parametrization of the
octupole shape \cite{Ro90,Ham91,HBXZ91}, we propose a collective
Hamiltonian which incorporates the interactions responsible for
the rotations associated with the different octupole
deformations. In addition we take into account the quadrupole
degrees of freedom and the appropriate higher order
quadrupole-octupole interaction. Below it will be shown that such
a general model Hamiltonian could incorporate the basic
properties of a nuclear system rotating under the above
assumption.

Although this assumption seems to be rather strong (since the
presence of stable octupole deformations in nuclei is not yet
a well elucidated problem) we suppose that it could give a natural
possibility to estimate the extent to which some of the observed
nuclear octupole bands carry the characteristics of the stable
octupole shapes. Generally, the proposed consideration will
provide a direct physical insight into the nuclear collective
motion as far as the shape of the system and the respective
moments of inertia are slightly changed under the collective
motion. Similarly to the case of the pure quadrupole
deformations this requirement will be naturally  satisfied for the
low angular momentum region of the spectrum which is, from another
perspective, accessible for detailed microscopic analysis, the
length of this region depending on the particular system.

Furthermore, we expect that in the higher angular momentum regions
the approach suggested will outline some general properties of the
system and thus will provide a relevant guide for respective more
detailed studies from both microscopic and phenomenological points
of view. In particular it will be shown that the model
formalism developed in the present work proposes a schematic
explanation of the recently observed \cite{DBoct00} ``beat''
patterns for the odd--even staggering in octupole bands of light
actinides based on recent experimental data
\cite{Cocks97,Cocks99}.

In Section 2 of the present work the octupole terms of the
Hamiltonian, classified by the irreducible representations
(irreps) of the octahedron group, will be described, while the
quadrupole terms and the octupole--quadrupole interaction will be
examined in Section 3, along with the bandhead term of the
Hamiltonian and the minimization procedure, which is a basic
ingredient of the present work. In Section 4 the diagonal parts
of the Hamiltonian will be analyzed and used for the production
of schematic odd--even staggering patterns, while the same
procedure will be repeated including the non-diagonal parts of
the Hamiltonian in Section 5. In Section 6 an analysis of the
model formalism under some restrictions on the permitted values of
the angular momentum projection on the body-fixed $z$-axis
will be presented, while further tests of the formalism will be performed 
in Section 7. Finally, Section 8 will contain discussion
of the present results, while in Section 9 a summary of the
present results and plans for future work will be given.

\section{Parametrization of the octupole deformation}

Our model formalism is based on the principle that the collective
properties of a physical system in which octupole correlations
take place can be expressed by the following most general
octupole field in the intrinsic (body-fixed) frame \cite{Ham91}
\begin{equation}
V_{3}=\sum_{\mu =-3}^{3}
\alpha_{3\,\mu}^{fix}Y^{*}_{3\,\mu} \ ,
\label{field1}
\end{equation}
with
$(\alpha_{3\,\mu}^{fix})^{*}=(-1)^{\mu}\alpha_{3\, -\mu}^{fix}$

This field can be written in the form \cite{Ham91}
\begin{equation}
V_{3}=\epsilon_{0}A_{2}+\sum_{i=1}^{3}\epsilon_{1}(i)F_{1}(i)+
\sum_{i=1}^{3}\epsilon_{2}(i)F_{2}(i) \ ,
\label{field2}
\end{equation}
where the quantities \cite{Ham91}
\begin{eqnarray}
A_{2}&=&-\frac{i}{\sqrt{2}}(Y_{3\, 2}-Y_{3\, -2})
     =\frac{1}{r^{3}}\sqrt{\frac{105}{4\pi}}xyz \ , \\
F_{1}(1)&=&Y_{3\, 0}=\frac{1}{r^{3}}\sqrt{\frac{7}{4\pi}}
           z(z^{2}-\frac{3}{2}x^{2}-\frac{3}{2}y^{2}) \ , \\
F_{1}(2)&=&-\frac{1}{4}\sqrt{5}(Y_{3\, 3}-Y_{3\, -3})
+\frac{1}{4}\sqrt{3}(Y_{3\, 1}-Y_{3\, -1}) \nonumber \\
        &=&\frac{1}{r^{3}}\sqrt{\frac{7}{4\pi}}
           x(x^{2}-\frac{3}{2}y^{2}-\frac{3}{2}z^{2}) \ , \\
F_{1}(3)&=&-i\frac{1}{4}\sqrt{5}(Y_{3\, 3}+Y_{3\, -3})
-i\frac{1}{4}\sqrt{3}(Y_{3\, 1}+Y_{3\, -1}) \nonumber \\
        &=&\frac{1}{r^{3}}\sqrt{\frac{7}{4\pi}}
           y(y^{2}-\frac{3}{2}z^{2}-\frac{3}{2}x^{2})\ , \\
F_{2}(1)&=&\frac{1}{\sqrt{2}}(Y_{3\, 2}+Y_{3\, -2})
         =\frac{1}{r^{3}}\sqrt{\frac{105}{16\pi}}
          z(x^{2}-y^{2})\ , \\
F_{2}(2)&=&\frac{1}{4}\sqrt{3}(Y_{3\, 3}-Y_{3\, -3})
+\frac{1}{4}\sqrt{5}(Y_{3\, 1}-Y_{3\, -1}) \nonumber  \\
        &=&\frac{1}{r^{3}}\sqrt{\frac{105}{16\pi}}
          x(y^{2}-z^{2})\ , \\
F_{2}(3)&=&-i\frac{1}{4}\sqrt{3}(Y_{3\, 3}+Y_{3\, -3})
+i\frac{1}{4}\sqrt{5}(Y_{3\, 1}+Y_{3\, -1}) \nonumber  \\
        &=&\frac{1}{r^{3}}\sqrt{\frac{105}{16\pi}}
          y(z^{2}-x^{2}),
\end{eqnarray}
(with $r^{2}=x^{2}+y^{2}+z^{2}$)
belong to irreducible representations (irreps) of the octahedron
group $O$. In particular, the first quantity (Eq. (3)~) belongs to
the one-dimensional irrep $A_{2}$, while the next three quantities
(Eqs (4)-(6)~) belong to the three-dimensional irrep  $F_{1}$ and
the last three quantities (Eqs (7)-(9)~) belong to the
three-dimensional irrep $F_{2}$. The seven real parameters
$\epsilon_{0}$ and $\epsilon_{r}(i)$ ($r=1,2;\, i=1,2,3$),
appearing in Eq. (2), determine the amplitudes of the various
components of the octupole deformation. Their relation to the
coefficients $\alpha_{3\,\mu}^{fix}$ has been given in Ref.
\cite{Ham91}.

As we have already seen, the quantities appearing in Eqs (3)-(9),
when expressed in terms of the cartesian coordinates $x$, $y$, and
$z$, contain linear combinations of terms cubic in the cartesian
variables. These specific linear combinations correspond to various
octupole shapes (as seen from their expressions in terms of the
spherical harmonics), and in addition correspond to the above
mentioned irreps of the octahedron group. Our proposition is the
following:

a) We construct a Hamiltonian using the same cubic terms appearing
in Eqs (3)-(9), but replacing the cartesian coordinates $x$, $y$,
$z$ by the angular momentum operators $\hat I_x$, $\hat I_y$, $\hat
I_z$ (with $\hat I^2 = \hat I_x^2 + \hat I_y^2 + \hat I_z^2$). For
example, the term $z^3$ is replaced by $\hat I_z^3$.

b) Due to the fact that the operators $\hat I_x$, $\hat I_y$, $\hat
I_z$ do not commute, while the coordinates $x$, $y$, $z$ do
commute, when making step a) we symmetrize each cubic expression
when we write it in terms of the angular momentum operators. For
example, the term $z (x^2-y^2)$ is replaced by $\hat I_z (\hat
I_x^2 - \hat I_y^2) + (\hat I_x^2-\hat I_y^2) I_z$.

c) During the procedure described above, the $r^3$ factors
appearing in the denominators of Eqs (3)-(9) are replaced by $\hat
I^3$ factors. In the final result we normalize with respect to
$\hat I^3$, i.e. we multiply the results by $\hat I^3$, an
operation which is equivalent to the transition to a unit sphere, a
natural thing to do since we are interested in surface shapes.
Indeed, this operation is equivalent to the multiplication of the
quantities appearing in Eqs (3)-(9) by $r^3$, an action which
eliminates their radial dependence, leading to a transition to the
unit sphere. The transition to the unit sphere is the reason that
the standard quadrupole and octupole operators are defined to be
proportional to $r^2 Y_{2\mu}$ and $r^3 Y_{3\mu}$ respectively
\cite{BM75}.

d) As a result, we obtain a Hamiltonian (as a function of the
angular momentum operators $\hat I_x$, $\hat I_y$, $\hat I_z$) the
terms of which correspond to the same octupole shapes which appear
in the Hamiltonian of Eqs (2)-(9).

e) The terms of the resulting Hamiltonian in addition belong to the
same irreps of the octahedron group as the terms of the original
Hamiltonian, appearing in Eqs (3)-(9). In other words, through this
procedure we determine the octahedron point symmetry properties of
the system in angular momentum space.

Our proposition is similar to the procedure used in Ref.
\cite{LuoLi} for the hexadecapole field.

The following Hamiltonian is then obtained:
\begin{equation}
\hat{H}_{oct}=\hat{H}_{A_{2}}+
\sum_{r=1}^{2}\sum_{i=1}^{3}\hat{H}_{F_{r}(i)} \ ,
\label{Hoctgen}
\end{equation}
with
\begin{eqnarray}
\hat{H}_{A_{2}}&=&{a}_{2}\frac{1}{4}
[(\hat{I}_x\hat{I}_y+\hat{I}_y\hat{I}_x)\hat{I}_z+
\hat{I}_z(\hat{I}_x\hat{I}_y+\hat{I}_y\hat{I}_x)] \ ,
\label{HA} \\
\hat{H}_{F_{1}(1)}&=&\frac{1}{2}{f}_{11}
\hat{I}_z(5\hat{I}_z^{2}-3\hat{I}^{2}) \ ,
\label{HF11} \\
\hat{H}_{F_{1}(2)}&=&\frac{1}{2}{f}_{12}
(5\hat{I}_x^{3}-3\hat{I}_x\hat{I}^{2}) \ ,
\label{HF12} \\
\hat{H}_{F_{1}(3)}&=&\frac{1}{2}{f}_{13}
(5\hat{I}_y^{3}-3\hat{I}_y\hat{I}^{2}) \ ,
\label{HF13} \\
\hat{H}_{F_{2}(1)}&=&{f}_{21}\frac{1}{2}
[\hat{I}_z(\hat{I}_x^{2}-\hat{I}_y^{2})+
(\hat{I}_x^{2}-\hat{I}_y^{2})\hat{I}_z] \ ,
\label{HF21} \\
\hat{H}_{F_2(2)}&=&{f}_{22}
(\hat{I}_x\hat{I}^{2}-\hat{I}_x^{3}-
\hat{I}_x\hat{I}_z^{2}-\hat{I}_z^{2}\hat{I}_x) \ ,
\label{HF22} \\
\hat{H}_{F_2(3)}&=&{f}_{23}
(\hat{I}_y\hat{I}_z^{2}+\hat{I}_z^{2}\hat{I}_y+
\hat{I}_y^{3}-\hat{I}_y\hat{I}^{2})
\label{HF23}
\end{eqnarray}
The Hamiltonian parameters ${a}_{2}$ and ${f}_{r\, i}$ ($r=1$ ,2;
$i=1$, 2, 3), appearing in Eqs (11)-(17), are formally related to
the parameters in Eq. (\ref{field2}) as follows
\begin{equation}
{a}_{2}=\epsilon_{0}\sqrt{\frac{105}{4\pi}}, \qquad
{f}_{1\, i}=\epsilon_{1}(i)\sqrt{\frac{7}{4\pi}}, \qquad
{f}_{2\, i}=\epsilon_{2}(i)\sqrt{\frac{105}{16\pi}},
\qquad \ i=1,2,3 .
\end{equation}

The non-vanishing matrix elements of the operators of Eqs
(\ref{HA})--(\ref{HF23}) in the states with collective angular
momentum $I$  are given in the Appendix.

In the Appendix we remark that the operator $\hat{H}_{F_{1}(1)}$
(Eq.~(\ref{HF11})~), which corresponds to the term $Y_{3\, 0}$
(see Eq. (4)~), characterized by axial deformation,  is the only
octupole operator possessing diagonal matrix elements. Below it
will be seen that this operator is of major importance in
determining the fine structure of collective bands with octupole
correlations. This fact does not come as a surprise, since it is
well known that the $Y_{3\, 0}$ (axial) deformation is the leading
mode in systems with reflection asymmetric shapes (see Ref.
\cite{BN96} for a relevant review).

\section{Inclusion of quadrupole degrees of freedom}

It is known, however,  that the use of the pure octupole field of
Eq. (\ref{field1}) is not sufficient for the description of the
collective properties of nuclei exhibiting octupole deformation,
since the quadrupole deformation is also present. Therefore one has
to consider the octupole degrees of freedom together with the
quadrupole deformations, and in addition one has to deal with their
coupling. A general treatment of a combined quadrupole--octupole
field has been given earlier in the framework of a general
collective model for coupled multipole surface modes
\cite{Ro88,Ro82}.

According to the above considerations it is reasonable to suggest
that the most general collective Hamiltonian of a system exhibiting
octupole deformations should also contain the standard (axial)
quadrupole rotation part
\begin{equation}
\hat{H}_{rot}= A\hat{I}^{2}+A'\hat{I}_{z}^{2} \ ,
\label{Hrot}
\end{equation}
where $A$ and $A'$ are the inertial parameters. In addition, it is
reasonable to try to describe the coupling between the quadrupole
and the octupole degrees of freedom by the following higher order
diagonal quadrupole--octupole interaction term, which corresponds
to the product $Y_{2\, 0}\ Y_{3\, 0}$,
\begin{equation}
\hat{H}_{qoc}=f_{qoc}\frac{1}{I^{2}}(15\hat{I}_{z}^{5}-
                  14\hat{I}_{z}^{3}\hat{I}^{2}+
                  3\hat{I}_{z}\hat{I}^{4}) \ ,
\label{Hqoc}
\end{equation}
since it is known that the axial deformation corresponding to
$Y_{3\, 0}$ is the leading mode in systems with reflection
asymmetric shapes \cite{BN96} and, in an analogous way, the axial
deformation corresponding to $Y_{2\, 0}$ is the leading mode in
systems with quadrupole deformations. The operator of Eq. (20) is
also normalized with respect to the multiplication factor $I^3$, in
the same way as the operators of Eqs (11)-(17) are, as described in
the previous Section. In other words, we use the product $I^3\
Y_{2\, 0}\ Y_{3\, 0}$, in order  to ensure that  all non-quadrupole
Hamiltonian terms will be of the same order.

As a result, the total Hamiltonian of the system can be written in
the form
\begin{equation}
\hat{H}=\hat{H}_{bh}+\hat{H}_{rot}+\hat{H}_{oct}+
\hat{H}_{qoc} \ ,
\label{Hgen}
\end{equation}
where
\begin{equation}
\hat{H}_{bh}=\hat{H}_{0}+f_{k}\hat{I}_{z} \ ,
\end{equation}
is a pure phenomenological part introduced in order to reproduce
the bandhead energy in the form
\begin{equation}
E_{bh}=E_{0}+f_{k}K \ ,
\end{equation}
were $E_{0}$ and $f_{k}$ are free parameters. The bandhead energy
$E_{bh}$ is considered proportional to $K$ in analogy to what
happens in the five-dimensional quadrupole oscillator model (see
Appendix 6B of Ref. \cite{BM75}), as well as in the standard
rotation--vibration model (see section 6.5 of Ref. \cite{GM96}).
The $K$-dependence of $E_{bh}$ plays an important role in our
approach, since it provides the correct value of the bandhead
angular momentum projection $K$ in the variation procedure
described below.

We remark that the Hamiltonian of Eq. (\ref{Hgen}) is not a
rotational invariant in general. It does not commute with the
total angular momentum operators and, as a result, any state with
given angular momentum $I$ is energy split with respect to the
quantum number $K$. Therefore, the physical relevance of this
Hamiltonian depends on the possibility to determine in a unique
way the angular momentum projection. Our basic assumption is that
$K$ is not frozen within the sequence of states of the collective
rotational band. We suggest that for any given angular momentum
$I$ the projection $K$ should be determined so as to minimize the
respective collective energy. The resulting octupole band is then
the yrast sequence of the energy levels produced by our model
Hamiltonian. It should be mentioned that a similar procedure has
been used in  Refs \cite{HM,MQ} in relation to the $\Delta I=2$
staggering effect in superdeformed nuclei.

\section{The diagonal parts of the Hamiltonian}

As a first step in testing our Hamiltonian we consider its diagonal
part
\begin{equation}
\hat{H}^{d}=\hat{H}_{bh}+\hat{H}_{rot}+\hat{H}_{oct}^{d}+
\hat{H}_{qoc} \ ,
\label{Hdiag}
\end{equation}
where the operator $\hat{H}_{oct}^{d}\equiv \hat{H}_{F_{1}(1)}$
represents the diagonal part of the pure octupole Hamiltonian
$\hat{H}_{oct}$ of Eq.~(\ref{Hoctgen}).

The following diagonal matrix element is then obtained
\begin{eqnarray}
E_{K}(I)&=&E_{0}+f_{k}K+AI(I+1)+A'K^{2}+
f_{11}\left( \frac{5}{2}K^{3}-\frac{3}{2}KI(I+1)\right)
\nonumber \\
&+&f_{qoc}\frac{1}{I^{2}}
\left( 15K^{5}-14K^{3}I(I+1)+
3KI^{2}(I+1)^{2}\right) \ .
\label{EKI}
\end{eqnarray}

Following the above mentioned assumption for the angular momentum
projection $K$, we determine the yrast sequence $E(I)$ by
minimizing for each value of $I$ the expression of Eq.~(\ref{EKI})
as a function of integer $K$ in the range $-I\leq K\leq I$. The
obtained energy spectrum depends on six model parameters: $E_{0}$
is a constant contributing to the bandhead energy; $f_{k}$ is
determined in order to correspond to the correct bandhead energy of
the octupole band for $I=1$, in which case one has $K=1$ as well
(This is appropriately modified in the case of bands starting with
higher $K$. For example in $K=5$ bands the lowest angular momentum
is $I=5$, and $f_k$ is fixed in order to provide the correct
bandhead energy for $I=K=5$.); $A$ and $A'$ are the quadrupole
inertial parameters which should in general correspond to the known
quadrupole shape (axes ratio) of the relevant nucleus; $f_{11}$ and
$f_{qoc}$ are the parameters of the diagonal octupole
(Eq.~(\ref{HF11})~) and quadrupole-octupole (Eq.~(\ref{Hqoc})~)
interactions respectively. In what follows we give fixed values to
the first four parameters ($E_0$, $f_k$, $A$, $A'$) and vary the
last two parameters ($f_{11}$, $f_{qoc}$) in order to examine the
influence of the last two terms of Eq. (25) on the odd--even
staggering pattern.

A schematic energy spectrum of this kind, obtained for a sample
set of parameter values, is given in Table 1. It is seen that the
``yrast'' values of the quantum number $K$, resulting from the
minimization procedure described above, gradually increase with
the increase of the angular momentum $I$. We remark that these
values of $K$  correspond to the local minima of the energy
expression of Eq.~(\ref{EKI}) as a function of $K$, which is
illustrated in Fig.~1 for the same set of parameters. We see that
these minima appear for positive values of $K$, are well
determined,  and their depth increases with the increase of the
angular momentum $I$.

These results indicate that the rotational motion of a system with
stable quadropole--octupole deformation is associated with complex
angular momentum dynamics, due to the complex shape contributions
to collectivity. More precisely, the following theoretical
proposition is formed: For this kind of rotational motion the
increase in the total angular momentum of the system is associated
with an energetically favorable increase in its third projection
$K$ on the body fixed axis. This means that the vector of the
angular momentum deviates ``step by step'' from the ``$x$-$y$''
body-fixed plane (which is perpendicular to the body-fixed
``z''-axis) as its magnitude increases. In other words, the higher
angular momenta will be attended by stronger precession, or
``wobbling motion'' of the system. These findings resemble the
recent results \cite{Ansari} obtained within the framework of a
self-consistent cranked HFB approach. Application of the method to
some Os isotopes has demonstrated that high-$K$ bands become
important in the high angular momentum region \cite{Ansari}.

Such a behavior of the spectrum could also be interpreted as a
multiband-crossing phenomenon, since the obtained yrast sequence
can be considered as the envelope of the curves with different
values of the quantum number $K$, as it is illustrated in Fig.~2.
This interpretation can, however, be considered as a mathematical
description, not necessarily implying the physical existence of
several bands with different values of $K$.

Thus, our schematic consideration suggests that the rotational
motion of a system with stable quadrupole--octupole deformations
invokes $K$-values essentially higher than the ones usually
considered in microscopic studies. We remark that the situation
considered here is essentially different from the case of pure
quadrupole rotations, where the bandhead $K$-value always provides
the minimum of the collective rotational energy along the whole
band (for example $E_{rot}=A I(I+1) + A'K^2$). Thus the obtained
wobbling motion appears as an effect of the higher multipolarity of
the considered collective interactions. The extent to which the
$K$-values appearing in our procedure provide physically reasonable
interpretations of various nuclear collective modes will be
considered in Section 6, where the limitation to $K\leq 3$ will be
studied.

Now we can examine the fine structure of the collective bands
obtained in the above model procedure. We see in Table 1 that the
$K$-values characterizing the levels with increasing $I$ are
grouped in couples containing two successive values of $I$ (one
even and one odd). This fact implies the presence of an odd--even
staggering effect, as we shall immediately see. A measure of the
odd--even staggering effect is the quantity
\begin{equation}
Stg(I)= 6\Delta E(I)-4\Delta E(I-1)-4\Delta E(I+1)+
\Delta E(I+2)+\Delta E(I-2)\ ,
\label{stag}
\end{equation}
with
\begin{equation}
\Delta E(I)=E(I+1)-E(I).
\end{equation}
The quantity $Stg(I)$ is proportional to
the discrete approximation of the fourth
derivative of the function $\Delta E(I)$, i.e.
proportional to the fifth derivative
of the energy $E(I)$, and is able to demonstrate fine deviations
from the pure rotational behavior, as it has been demonstrated in
Ref. \cite{DBoct00} for the octupole bands of several light
actinides. An analog of this quantity has been introduced earlier
\cite{Fli1,Fli2,Ced} for the study of $\Delta I=2$ staggering in
nuclear superdeformed bands.

The way in which the appearance of the $K$-values in couples for
successive values of $I$ leads to odd--even staggering can be seen
from Table 1, using the second term of Eq. (25) as an example. In
Table 1 we see that couples of $K$ values appear for $I\geq 8$. In
this region the contribution of the second term of Eq. (25) to
$\Delta E(I)$ is zero for even values of $I$, while it is $+f_k$
for odd values of $I$. Then we easily see from Eq. (26) that the
contribution of the second term of Eq. (25) to $Stg(I)$ is $+8 f_k$
for odd values of $I$, while it is $-8 f_k$ for even values of $I$.
Therefore the second term of Eq. (25) leads to odd--even staggering
of constant amplitude. The 4th, 5th and 6th terms of Eq. (25)
depend nonlinearly on $K$ and/or $I$, and therefore give odd--even
staggering of varying amplitude. In the case of the 4th term, it is
easy to see that it gives staggering with amplitude increasing
linearly as a function of I. (The terms $E_0$ and $A I(I+1)$ do not
contribute to the odd--even staggering, as it is known from Ref.
\cite{DBoct00}.)

Several examples of odd--even staggering occurring from Eq. (26) for
different sets of parameter values used in Eq. (25) are given in
Fig. 3. The parameters $E_0$ and $f_k$ have been kept constant in
all parts of Fig. 3, while the parameters $A$ and $A'$ have been
kept constant in all parts of Fig. 3 except the last one (Fig.
3(f)~). In contrast, the parameters $f_{11}$ and $f_{qoc}$ have
been given different values in the various parts of Fig. 3.

Fig.~3(a) illustrates a long odd--even staggering pattern which
looks similar to the ``beats'' observed in the octupole bands of
some light actinides \cite{DBoct00} (with $^{220}$Ra, $^{224}$Ra
and $^{226}$Ra being probably the best examples), a difference
being that in the realistic cases of Ref. \cite{DBoct00} the
amplitude of the ``beats'' seems to be decreasing with increasing
$I$, while in Fig. 3(a) the opposite seems to hold.

In Fig.~3(b) the increased values of $f_{11}$ and $f_{qoc}$
provide a wide angular momentum region (up to $I\sim 40$) with a
regular staggering pattern.

The further increase of $f_{qoc}$ results
in a staggering pattern with different amplitudes,
shown in Fig.~3(c).

The further increase of $f_{11}$ and $f_{qoc}$ leads to a
staggering pattern with many ``beats'', as shown in Fig~3(d).
Notice that in Fig. 3(d) the first three ``beats'' are completed by
$I\approx 40$, while in Fig. 3(a) the first three ``beats'' are
completed by $I\approx 70$.

Fig.~3(e) illustrates what happens in the case of vanishing
quadrupole--octupole interaction term ($f_{qoc}=0$), keeping the
rest of the parameters the same as in Figs 3(b) and 3(c). We see
that in the present case the ``beat'' effect occurs very
frequently, while in Figs 3(b) and 3(c) the change was much slower.

Finally, an example with almost constant staggering amplitude is
shown in Fig.~3(f). It resembles the form of the odd--even
staggering predicted in the SU(3) limit of various algebraic models
(see Ref. \cite{DBoct00} for details and relevant references). It
also resembles the odd--even staggering seen in some octupole bands
of light actinides \cite{DBoct00} (with $^{222}$Rn being probably
the best example).

\section{The non-diagonal parts of the Hamiltonian}

After seeing in the previous section the main features of the
diagonal part of the Hamiltonian, we can now focus our attention
on the general Hamiltonian of Eq. (\ref{Hgen}), including the
various non-diagonal terms given in Eqs (\ref{HA}),
(\ref{HF12})--(\ref{HF23}). The main problem in this case is the
fact that $K$ is in general not a good quantum number. We can,
however, make analysis for small values of the respective
parameters ($a_2$, $f_{12}$, $f_{13}$, $f_{21}$, $f_{22}$,
$f_{23}$), which keep $K$ ``asymptotically'' good. Keeping these
parameters small means that we use a weak $K$-bandmixing
interaction, which guarantees that for any explicit energy minimum
appearing in the diagonal case the corresponding eigenvalue of the
perturbed Hamiltonian will be uniquely determined. Thus we are
able to obtain for the perturbed Hamiltonian a $K$-mixed yrast
energy sequence analogous to the one we got in the previous
section for the diagonal part of the Hamiltonian.

Our numerical analysis of the Hamiltonian eigenvector systems
shows that the parameters of the non-diagonal terms should be
smaller by an order of magnitude in comparison to the value of
the parameter $f_{11}$. In addition, by examining the
corresponding matrix elements in the Appendix, we deduce that the
following pairs of non-diagonal terms give identical contributions
to the energy spectrum: $\hat{H}_{A_{2}}$ and
$\hat{H}_{F_{2}(1)}$; $\hat{H}_{F_{1}(2)}$ and
$\hat{H}_{F_{1}(3)}$; $\hat{H}_{F_{2}(2)}$ and
$\hat{H}_{F_{2}(3)}$. Therefore it suffices to keep from now on
only the terms $\hat{H}_{F_1(2)}$, $\hat{H}_{F_2(1)}$,
$\hat{H}_{F_2(2)}$.

In Fig.~4 two staggering patterns in the presence of $K$-bandmixing
terms are illustrated. In the calculation we have included, as we
have already mentioned, the non-diagonal terms
$\hat{H}_{F_{1}(2)}$, $\hat{H}_{F_{2}(1)}$, and
$\hat{H}_{F_{2}(2)}$, along with the already considered diagonal
Hamiltonian of Eq. (\ref{Hdiag}). The three non-diagonal terms have
been included with two different sets of parameters (both obeying
the above mentioned condition that the values of the parameters
accompanying the non-diagonal terms should be at least an order of
magnitude smaller than the value of the parameter $f_{11}$), shown
in Figs 4(a) and 4(b), while the parameters of the diagonal part
have been kept the same as in Fig.~3(b) (and in Table~1). Comparing
Figs 3(b) and 4(a) we see that the non-diagonal terms affect more
severely the higher angular momentum region by decreasing the
staggering amplitude. Moreover, the larger values for the
parameters of the non-diagonal terms, shown in Fig.~4(b), reduce
more seriously the staggering pattern in the higher angular
momentum region, as one can see by comparing Figs 3(b) and 4(b) and
noticing the different scales on the vertical axes of the two
figures. The pattern appearing in Fig. 4(b) resembles the
experimental situation in $^{218}$Rn and $^{228}$Th \cite{DBoct00}
(odd--even staggering with amplitude decreasing as a function of
$I$).

\section{Low-$K$ ($K\leq 3$) analysis of the model formalism}

A discussion on the $K$-values appearing in our model
consideration is appropriate in this place.

Usually from microscopic point of view the eigenvalues of the total
angular momentum projection $I_z$ (which, in case of axial
symmetry, coincides with the intrinsic momentum projection $J_z$)
are restricted to $K=\leq 3$. This requirement is well justified in
the case of octupole vibrations of quadrupole deformed nuclei. It
reflects the shell model concept that the octupole degrees of
freedom could be generated by an octupole-octupole $\lambda =3$
interaction (where $\lambda$ stands, as usual, for the
multipolarity) which couples single-particle states of opposite
parity \cite{BN96}. In particular, this assumption gives a
satisfactory interpretation for the fine structure of negative
parity rotational bands built on octupole vibrations.

In view of the above,  it is interesting to examine our
formalism in the case when the space of the $K$-values is
restricted to $K_{max}=3$. Indeed such a test is important
especially for the low angular momentum regions, where the
intrinsic structure of octupole degrees of freedom is
microscopically well studied.

Along these lines, we performed a schematic
calculation including the values
$K=0$, 1, 2, 3 in the angular momentum region up to
$I=10$, as well as another calculation including the values
$K=0$, 1, 2, 3 in the angular momentum region up to
$I=16$. The yrast sequences obtained are given in Tables 2 and 3
respectively, while the resulting staggering patterns are shown in
Fig. 5. We see that our formalism provides a regular staggering
pattern for the states up to $I=10$ [Fig. 5(a)] and a ``beat''
pattern for the spectrum up to $I=16$ [Fig. 5(b)]. These examples
(obtained for the parameter values given in Tables 2 and 3,
respectively) illustrate that our approach is capable to provide
reasonable staggering patterns under the restriction to $K\leq 3$,
at least in the low angular momentum region with $I=10$-16. In
other words, the schematic model gives reasonable results under the
restriction to $K\leq 3$, at least in the microscopically
accessible angular momentum regions.

It is, however, important to remark that our approach suggests a
rather more extended treatment of the collective octupole degrees
of freedom, which is expected to be useful in view of the
increasing bulk of data indicating possible stable octupole
deformations in several nuclei, the light actinides providing the
best examples. Some of these bands (the octupole bands in
$^{224-226}$Ra and $^{224-228}$Th, for example) possess rather well
pronounced rotational structures (see Table~1 in
Ref.~\cite{DBoct00}). Our proposition is that in these cases the
concept of a rotating quadrupole--octupole shape should be more
evident. In these cases, in contrast to the pure octupole
vibrations, it is not {\it a priori} clear why the combined
quadrupole--octupole intrinsic configurations should be restricted
to $K \leq 3$ projection values. Namely for these kinds of
rotational motion our collective formalism suggests that $K$ could
be higher than $3$. One can easily see that for the relatively well
deformed nuclei, such as $^{224-226}$Ra and $^{224-228}$Th, the
Nilsson orbitals do not exclude the presence of reasonable $K>3$
intrinsic configurations \cite{Quentin}. This remark is of
particular importance for the higher angular momentum region ($I
\geq 10$-16), where pair breaking effects are possible.

Another important question to be examined is whether the energy
gain in case of yrast rotations with $K>3$ is larger than the
intrinsic energy necessary to raise $K$ above $3$. There is no {\it
a priori} answer to this question, which would be an interesting
problem for future combined efforts from both collective and
microscopic points of view. Here we just remark that for the
collective motion considered the energy gain rapidly increases with
the increase of the angular momentum (see the depth of the minima
in the curves of Fig.~1).

\section{Further tests of the model formalism} 

The restriction  to  $K \leq 3$ values, used in the previous section, 
demonstrated that odd--even staggering and ``beat'' patterns 
can indeed appear in this special case. This particular collection 
of $K$ values ($K=0$, 1, 2, 3) has certain probability to appear 
low in energy in the case of nuclei exhibiting octupole deformation, 
but other possibilities also exist. In fact, it is well known from 
experiment that the various $K$ values are built up through the 
pair-breaking mechanism. As a result, $K$ is not a smoothly increasing 
function (while $I$ is). Which $K$ values will appear lower in energy 
in each nucleus depends strongly on the microscopic structure of the 
nucleus. One is therefore confronted with the following questions:
 
1) Given a collection of $K$ values, which appear low in energy in a specific 
nucleus because of microscopic reasons, is the present formalism 
predicting odd--even staggering or not?    

2) What is the meaning of minimization with respect to $K$ in this case?

3) In which order do the various allowed values of $K$ appear as $I$ is 
increasing? 

We have performed several numerical tests, which need not to be reproduced 
here in detail, in order to answer these questions. The following 
conclusions have been reached: 

1) Given a collection of $K$ values which are non-successive integers 
(for example, $K=5$, 7, 9, 10, 15), odd--even staggering and ``beat''
patterns do appear for certain values of the Hamiltonian parameters.
One is therefore persuaded that in principle the model formalism can produce 
``beat'' patterns for any microscopically directed collection of $K$ 
values.  

2) In the just mentioned case, $K$ is not a continuous variable. However,
one can still minimize the energy with respect to $K$, in the following 
sense: Given a collection of $K$ values, for each value of $I$ one 
determines (among the members of the collection) the $K$ value for which 
the energy is minimum. 

3) In all cases the allowed values of $K$ appear in increasing order 
as $I$ is increasing, similarly to what occurs in Tables 1--3. 

In all of the above we have been assuming that $K$ takes only integer 
values, i.e. that $K$ is a good quantum number. This is an assumption 
of the model, which is quite usual in the realm of phenomenological 
models (but not in the case of microscopic models), in which $K$ is assumed 
to be a good (or asymptotically good) quantum number. One might 
wonder, however, how strongly the appearance of odd--even staggering 
and ``beat'' patterns in the present model depends on this assumption.  
In order to confront this question, we have performed the following 
tests: We repeated the calculations concerning Figs. 3(a) and 3(b) 
allowing $K$ to vary not with a step of 1.0, but with a step of 0.5,
or 0.1, or 0.01, or 0.0001, as well as allowing $K$ to be a continuous 
variable. The results need not be reproduced here in full detail, 
but the conclusion was that in both cases the odd--even staggering 
drops dramatically as the step of the variation of $K$ decreases, 
and practically vanishes when $K$ is a continuous variable. 
One therefore concludes that the appearence of odd--even staggering 
and ``beat'' patterns in the present model strongly depends on the 
assumption that $K$ is a good (or asymptotically good) quantum number, 
taking integer values, an assumption which is rather common among 
phenomenological models. In the other hand, as it has already been 
mentioned in the beginning of this section, practically any
microscopically directed set of integer values of $K$ can lead 
to the appearance of odd--even staggering and ``beat'' patterns 
within the framework of the present model. 

In conclusion, the assumption that $K$ is a good (or asymptotically good)
quantum number, therefore taking integer values only, is a fundamental one 
as far as the appearance of odd--even staggering and ``beat'' patterns 
within the present model is concerned. Once $K$ is assuming only integer 
values, odd--even staggering and ``beat'' patterns can in principle 
be constructed (for appropriate parameter values in the Hamiltonian) 
for any set of microscopically directed $K$ values lying low in energy.  

\section{Discussion}

The staggering patterns illustrated so far (Figs~3 and 4) cover
almost all odd--even staggering patterns seen in nuclear octupole
bands. The amplitudes obtained for the sets of parameters
considered vary up to 300 keV. Staggering patterns with larger
values of $Stg(I)$ can be easily obtained for different parameter
sets. On this basis it is reasonable to assume that the model
parameters can be adjusted appropriately so as to reproduce the
staggering patterns seen in the octupole bands of the light
actinides \cite{DBoct00}, as well as in rotational negative parity
bands with $K\geq 1$ built on octupole vibrations. In the fitting
procedure one should take into account the minimization of the
energy with respect to $K$ for each given value of $I$. Work in
this direction is in progress.

At this point the following comments on the structure of the
collective interactions used and the related symmetries are in
place:

1) The above mentioned fact that the six non-diagonal terms of the
octupole Hamiltonian can be arranged into three pairs, the terms
belonging to the same pair giving equal contributions, indicates
that only four terms of the octupole Hamiltonian (the diagonal
term plus three appropriately chosen non-diagonal terms) suffice
for the determination of the energy spectrum. This result
reflects the fact that in the intrinsic frame of reference three
octupole degrees of freedom, from the total of seven ones, are
related to the orientation angles. In the present case we chose
in Sec. 5 and in Fig. 4 to keep the diagonal term
$\hat{H}_{F_{1}(1)}$ and the non-diagonal terms
$\hat{H}_{F_{1}(2)}$, $\hat{H}_{F_{2}(1)}$ and
$\hat{H}_{F_{2}(2)}$. Our analysis (related to the collective
rotations of the system) gives a natural way for determining the
four collective octupole interaction terms which give independent
contributions.

2) The symmetry of the various combinations of spherical harmonics
appearing in Eqs (3)-(9) has been considered in detail in Ref.
\cite{HBXZ91}. The $Y_{30}$ term is axially symmetric, i.e. it has
the symmetry of the D$_{\infty}$ group, while the term $Y_{3\, 1}
-Y_{3\, -1}$ has the symmetry of the C$_{2v}$ group, the term
$Y_{3\, 2}+Y_{3\, -2}$ has the symmetry of the T$_d$ group, and the
term $Y_{3\, 3}-Y_{3\, -3}$ has the symmetry of the D$_{3h}$ group.
>From the correspondence between Eqs (3)-(9) and Eqs (11)-(17) we
see that the diagonal term $\hat{H}_{F_1(1)}$ is axially symmetric,
while the non-diagonal terms $\hat{H}_{F_{1}(2)}$,
$\hat{H}_{F_{2}(1)}$ and $\hat{H}_{F_{2}(2)}$ (mentioned in the
previous item 1)~) are constructed by using the combinations
$(Y_{3\, 1}-Y_{3\, -1})$, $(Y_{3\, 2}+Y_{3\, -2})$, and $(Y_{3\, 3}
-Y_{3\, -3})$.

3) Concerning the role of the various terms in the odd--even
staggering effect, our schematic results of Secs. 4 and 5
indicate that the non-diagonal $K$-bandmixing interactions suppress
the staggering pattern, while the axially symmetric term
$\hat{H}_{F_{1}(1)}$ is able to provide a ``beat'' staggering
behavior (see Fig.~3(e)). The quadrupole--octupole term
$\hat{H}_{qoc}$ does influence the staggering pattern, providing
wider angular momentum regions with regular staggering. In short,
the axially symmetric term seems to be the most important one
(among the terms involving the octupole degrees of freedom) for the
production of ``beat'' behavior in the odd--even staggering
pattern. However, it should be remembered at this point that the
octupole and octupole--quadrupole terms are not the only ones
contributing to the odd--even staggering effect. As we have already
mentioned in Sec. 4, the term $f_k K$, coming from the bandhead
energy, makes to the odd--even staggering a contribution of
constant amplitude, while the $A' K^2$ term, coming from the
quadrupole part of the Hamiltonian, makes a contribution to the
odd--even staggering with an amplitude which is increasing linearly
as a function of $I$. It should be noticed, though, that these
remarks are based only on a few schematic calculations and in no
way are final conclusions. In order to reach safer conclusions
about the relative importance of the various terms in giving rise
to a ``beat'' behavior of the odd--even staggering, one has to
perform detailed fits to the octupole bands of the light actinides,
as mentioned above.

4) As has already been commented is Secs. 4 and 6 the most general
schematic results we have obtained suggest that in the high angular
momentum region some high $K$ values should be involved (as a
result of the minimization of the energy with respect to $K$ for
each given value of $I$, which is an important ingredient of our
approach). The fact that the $K$ quantum number is not a good
quantum number (not even approximately) of the relevant states has
been realized long ago \cite{NV70a,NV70b,V70}. In microscopic
calculations in the rare earth region (with $152 \leq A \leq 190$)
\cite{NV70a}, in which the values $K=0$, 1, 2, 3 have been
included, it has been seen that in the beginning of the region the
values $K=0$, 1 are important for the lowest $3^-$ state, while in
the middle of the region the values $K=1$, 2 are important and in
the far end of the region the values $K=2$, 3 are important
\cite{NV70a}. These results are consistent with our low-$K$ model
analysis given in Section 6. The same authors have dealt with the
actinide region ($A\geq 222$) in Ref. \cite{NV70b}. One of the
authors of Refs \cite{NV70a,NV70b} in Ref. \cite{V70} finds that
the restriction to $K\leq 3$ is not justifiable for large energies.
These findings are in agreement with our results given in  Table 1
as well as with our comments in the end of Sec. 6.

5) Concerning the analysis in Sec. 5 and the above comments 3) and
4), we remark that the restrictions imposed on the non-diagonal
Hamiltonian terms (keeping $K$ asymptotically good) are rather
strong and  reflect the particular physical assumptions of our
consideration. However, for a more general quantum mechanical
system there is no principal reason to restrict the analysis by
assuming the presence of small non-axial deformations only
\cite{Pittel}. In this direction an extension of the present work
not limited to small non-diagonal contributions could be of
interest.

\section{Conclusions and outlook}

In summary, we have considered a Hamiltonian involving octupole and
quadrupole terms, along with an octupole--quadrupole interaction
and a $K$-dependent bandhead term. The octupole terms have been
classified by using the irreducible representations (irreps) of the
octahedron group O. By minimizing the energy with respect to the
angular momentum projection $K$ for each given value of the angular
momentum $I$ we have reached the conclusion that $K$ is increasing
with increasing $I$, even in the case in which $K$ is allowed to assume 
only a resticted set of microscopically dictated values, a result which can be
interpreted as corresponding to a wobbling motion. Various terms of the
Hamiltonian give rise to odd--even staggering depending on $I$ in
different ways, in this way making the Hamiltonian able to produce
several different odd--even staggering patterns, some of which have
been observed in the octupole bands of light actinides. In order to
examine the relative importance of the various terms giving rise to
odd--even staggering, detailed fits of the octupole bands of the
light actinides should be performed, a procedure which is not very
simple because of the above mentioned minimization procedure
involved in it. In addition, the applicability of the present
formalism to rotational bands with $K\geq 1$ built on octupole
vibrations, which also demonstrate odd--even staggering effects,
should be examined. The present Hamiltonian, because of the variety
of odd--even staggering patterns it can produce, seems to be a good
starting point for systematizing the different odd--even staggering
patterns seen in octupole bands, as well as in rotational bands
built on octupole vibrations. Work in this direction is in
progress.

\bigskip \bigskip

\noindent {\bf Acknowledgments}

\medskip
The authors are grateful to Prof. P. Quentin for several
illuminating discussions at various stages of the development of
this work, as well as to Prof. S. Pittel for a careful reading of
manuscript and useful comments. One of authors (NM) is grateful to
Prof. N. Lo Iudice for hospitality during his stay in Universit\`a
di Napoli ``Federico II'' and for detailed discussions on the
subject of this work. This work has been supported by the Bulgarian
National Fund for Scientific Research under contract no
MU--F--02/98. Last but not least, the authors are grateful to an 
unknown referee, whose questions and suggestions led to the 
development of Sections 6 and 7 of the article. 
\medskip

This article is dedicated to the memory of Roussy Petkov Roussev,
dearest colleague and friend, who passed away soon after the
compilation of this work.

\newpage
\bigskip \bigskip
\noindent {\bf Appendix}

\medskip
Non-zero matrix elements of the operators of Eqs
(\ref{HA})--(\ref{HF23}) in the states $| I K \rangle$ with total
angular momentum $I$ and projection $K$, with $X=I(I+1)$:
\begin{eqnarray}
\left\langle I\, K+2|\hat{H}_{A_2}|I\, K\right\rangle&=&
\frac{i}{4}{a}_{2}(K+1)
\sqrt{X-(K+1)(K+2)}\sqrt{X-K(K+1)}, \nonumber \\
%
\left\langle I\, K-2|\hat{H}_{A_2}|I\, K\right\rangle&=&
-\frac{i}{4}{a}_{2}(K-1)
\sqrt{X-(K-1)(K-2)}\sqrt{X-K(K-1)}, \nonumber \\
%
\left\langle I\, K|\hat{H}_{F_{1}(1)}|I\, K\right\rangle&=&
\frac{1}{2}{f}_{11}K(5K^{2}-3X),  \nonumber \\
%
\left\langle I\, K+1|\hat{H}_{F_{1}(2)}|I\, K\right\rangle&=&
\frac{1}{16}{f}_{12}(3X-15K^{2}-15K-10)
\sqrt{X-K(K+1)}, \nonumber \\
%
\left\langle I\, K-1|\hat{H}_{F_{1}(2)}|I\, K\right\rangle&=&
\frac{1}{16}{f}_{12}(3X-15K^{2}+15K-10)
\sqrt{X-K(K-1)}, \nonumber \\
%
\left\langle I\, K+1|\hat{H}_{F_{1}(3)}I\, |K\right\rangle&=&
\frac{i}{16}{f}_{13}(3X-15K^{2}-15K-10)
\sqrt{X-K(K+1)}, \nonumber \\
%
\left\langle I\, K-1|\hat{H}_{F_{1}(3)}|I\, K\right\rangle&=&
-\frac{i}{16}{f}_{13}(3X-15K^{2}+15K-10)
\sqrt{X-K(K-1)},  \nonumber \\
%
\left\langle I\, K+2|\hat{H}_{F_{2}(1)}|I\, K\right\rangle&=&
\frac{1}{2}{f}_{21}(K+1)\sqrt{X-(K+1)(K+2)}
\sqrt{X-K(K+1)}, \nonumber \\
%
\left\langle I\, K-2|\hat{H}_{F_{2}(1)}|I\, K\right\rangle&=&
\frac{1}{2}{f}_{21}(K-1)\sqrt{X-(K-1)(K-2)}
\sqrt{X-K(K-1)}, \nonumber \\
%
\left\langle I\, K+1|\hat{H}_{F_{2}(2)}|I\, K\right\rangle&=&
\frac{1}{8}{f}_{22}(X-5K^{2}-5K-2)
\sqrt{X-K(K+1)}, \nonumber \\
%
\left\langle I\, K-1|\hat{H}_{F_{2}(2)}|I\, K\right\rangle&=&
\frac{1}{8}{f}_{22}(X-5K^{2}+5K-2)
\sqrt{X-K(K-1)},  \nonumber \\
%
\left\langle I\, K+1|\hat{H}_{F_{2}(3)}|I\, K\right\rangle&=&
-\frac{i}{8}{f}_{23}(X-5K^{2}-5K-2)
\sqrt{X-K(K+1)},  \nonumber \\
%
\left\langle I\, K-1|\hat{H}_{F_{2}(3)}|I\, K\right\rangle&=&
\frac{i}{8}{f}_{23}(X-5K^{2}+5K-2)
\sqrt{X-K(K-1)}.   \nonumber
\end{eqnarray}

\newpage

\begin{table}
\caption{The ``yrast'' energy levels, $E(I)$ (in keV), and the
respective $K$-values (in $\hbar$) obtained from
Eq.~(\protect\ref{EKI}) for the parameter set $E_{0}=500$ keV,
$f_{k}=-7.5$ keV, $A=12$ keV, $A'=6.6$ keV, $f_{11}=0.56$ keV,
$f_{qoc}=0.085$ keV, by minimizing the energy with respect to $K$
for each given value of $I$. See Sec. 4 for further discussion. }

    \bigskip

    \begin{center}
    \begin{tabular}{ccccccccc}
    \rule{0em}{2.2ex}
\\
\hline\hline
$I$&$E(I)$&$K$&$I$&$E(I)$&$K$&$I$&$E(I)$&K \\
\hline
1 & 522.772& 1&  13&  2335.81&  5 & 25&  5453.12&  11 \\
2 & 568.327& 1&  14&  2576.57&  6 & 26&  5694.49&  12 \\
3 & 637.095& 1&  15&  2827.57&  6 & 27&  5935.50&  12 \\
4 & 728.710& 1&  16&  3082.36&  7 & 28&  6157.50&  13  \\
5 & 840.857& 2&  17&  3344.94&  7 & 29&  6378.29&  13  \\
6 & 971.155& 2&  18&  3608.18&  8 & 30&  6575.37&  14  \\
7 & 1123.22& 2&  19&  3877.05&  8 & 31&  6770.62&  14  \\
8 & 1288.09& 3&  20&  4143.16&  9 & 32&  6937.23&  15  \\
9 & 1472.71& 3&  21&  4413.03&  9 & 33&  7101.62&  15  \\
10& 1668.56& 4&  22&  4676.45&  10& 34&  7232.21&  16  \\
11& 1880.56& 4&  23&  4942.01&  10& 35&  7360.44&  16  \\
12& 2101.68& 5&  24&  5197.18&  11& 36&  7449.45&  17  \\
    \hline\hline
    \end{tabular}
    \end{center}
    \label{tab:spec}
    \end{table}

\ \ \ \ \ \


\begin{table}
\caption{Same as Table 1, but for the parameter set $E_{0}=0$
keV, $f_{k}=0$ keV, $A=12$ keV, $A'=9.96$ keV, $f_{11}=0.75$ keV,
$f_{qoc}=0.15$ keV. The restriction $K \leq 3$ has been imposed.
See Sec. 6 for further discussion.  }

    \bigskip

    \begin{center}
    \begin{tabular}{ccccccc}
    \rule{0em}{2.2ex}
\\
\hline\hline
$I$&$E(I)$&$K$& &$I$&$E(I)$&$K$ \\
\hline
1 & 24.00&  0& & 6&   487.94&   1 \\
2 & 72.00&  0& & 7&   640.19&   2 \\
3 & 144.00& 0& & 8&   811.21&   2 \\
4 & 237.97& 1& & 9&   994.40&   3 \\
5 & 351.64& 1& & 10&  1194.18&  3 \\
    \hline\hline
    \end{tabular}
    \end{center}
    \label{tab:lowspec1}
    \end{table}

\ \ \ \ \ \


\begin{table}
\caption{Same as Table 1, but for the parameter set
$E_{0}=100$ keV, $f_{k}=0.896$ keV, $A=19.2$ keV, $A'=18.24$ keV,
$f_{11}=0.49$ keV, $f_{qoc}=0.074$ keV. The restriction $K\leq 3$
has been imposed. See Sec. 6 for further discussion.}

    \bigskip

    \begin{center}
    \begin{tabular}{ccccccc}
    \rule{0em}{2.2ex}
\\
\hline\hline
$I$&$E(I)$&$K$& &$I$&$E(I)$&$K$ \\
\hline
1 & 138.40& 0& & 9& 1803.42& 1 \\
2 & 215.20& 0& &10& 2177.43& 1 \\
3 & 330.40& 0& &11& 2580.53& 2 \\
4 & 484.00& 0& &12& 3017.25& 2 \\
5 & 676.00& 0& &13& 3490.32& 2 \\
6 & 905.66& 1& &14& 3991.13& 3 \\
7 &1167.54& 1& &15& 4521.73& 3 \\
8 &1466.79& 1& &16& 5087.69& 3 \\
    \hline\hline
    \end{tabular}
    \end{center}
    \label{tab:lowspec2}
    \end{table}

\ \ \ \ \ \
\newpage
    \begin{center}
    {\bf Figure Captions}
    \end{center}
    \bigskip\bigskip

\noindent
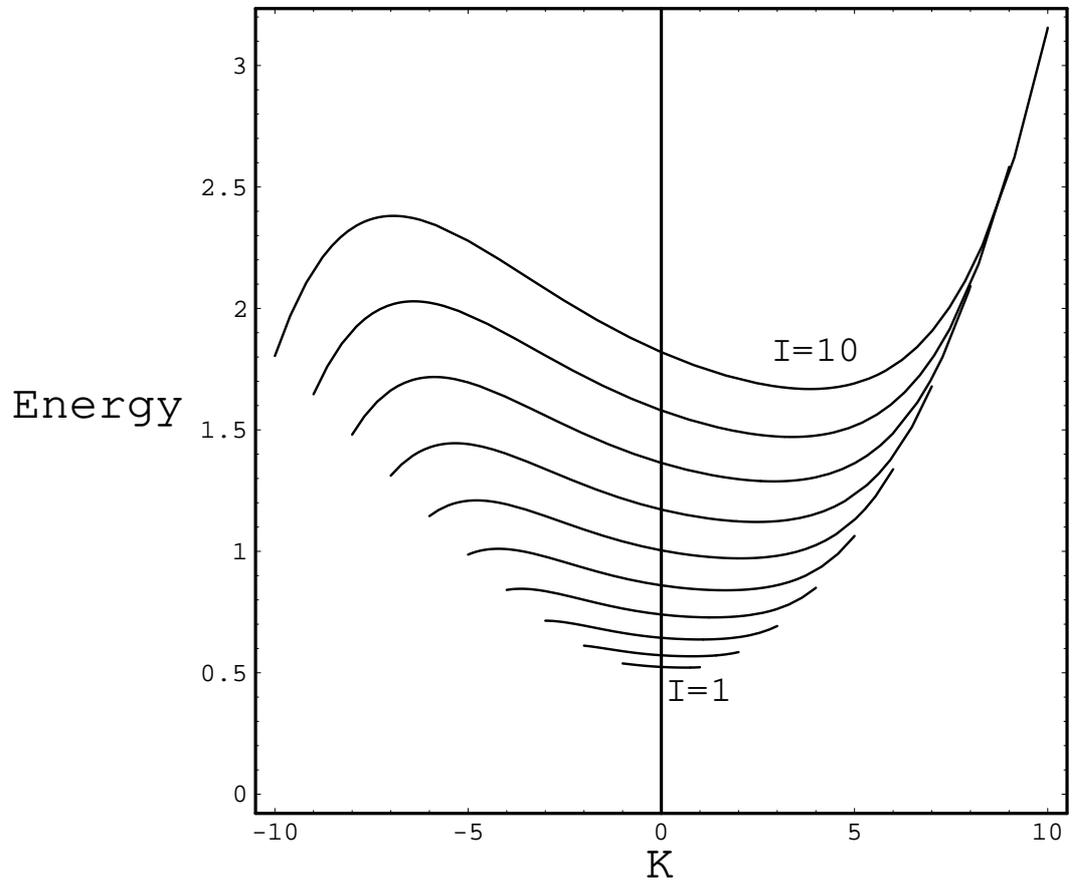
{\bf Figure 1.} The diagonal energy matrix element $E_{K}(I)$ (in
MeV), Eq.~(\protect\ref{EKI}), is plotted as a function of $K$ (in
$\hbar$) for $I=1,2,...,10$, for the parameter set $E_{0}=500$ keV,
$f_{k}=-7.5$ keV, $A=12$ keV, $A'=6.6$ keV, $f_{11}=0.56$ keV,
$f_{qoc}=0.085$ keV.
\bigskip

\noindent
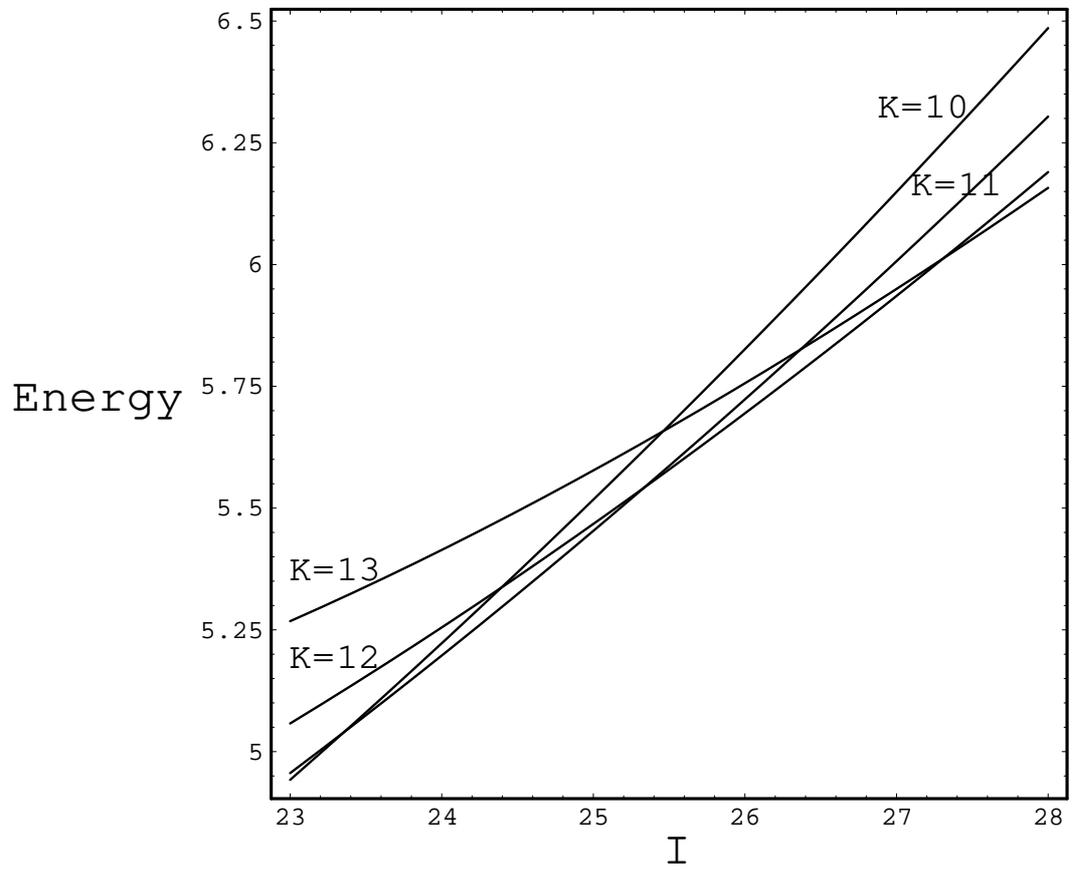
{\bf Figure 2.} The diagonal energy matrix element $E_{K}(I)$ (in
MeV), Eq.~(\protect\ref{EKI}), is plotted as a function of $I$ (in
$\hbar$) for $K=10,11,12,13$, for the parameter set of Figure 1.
\bigskip

\noindent
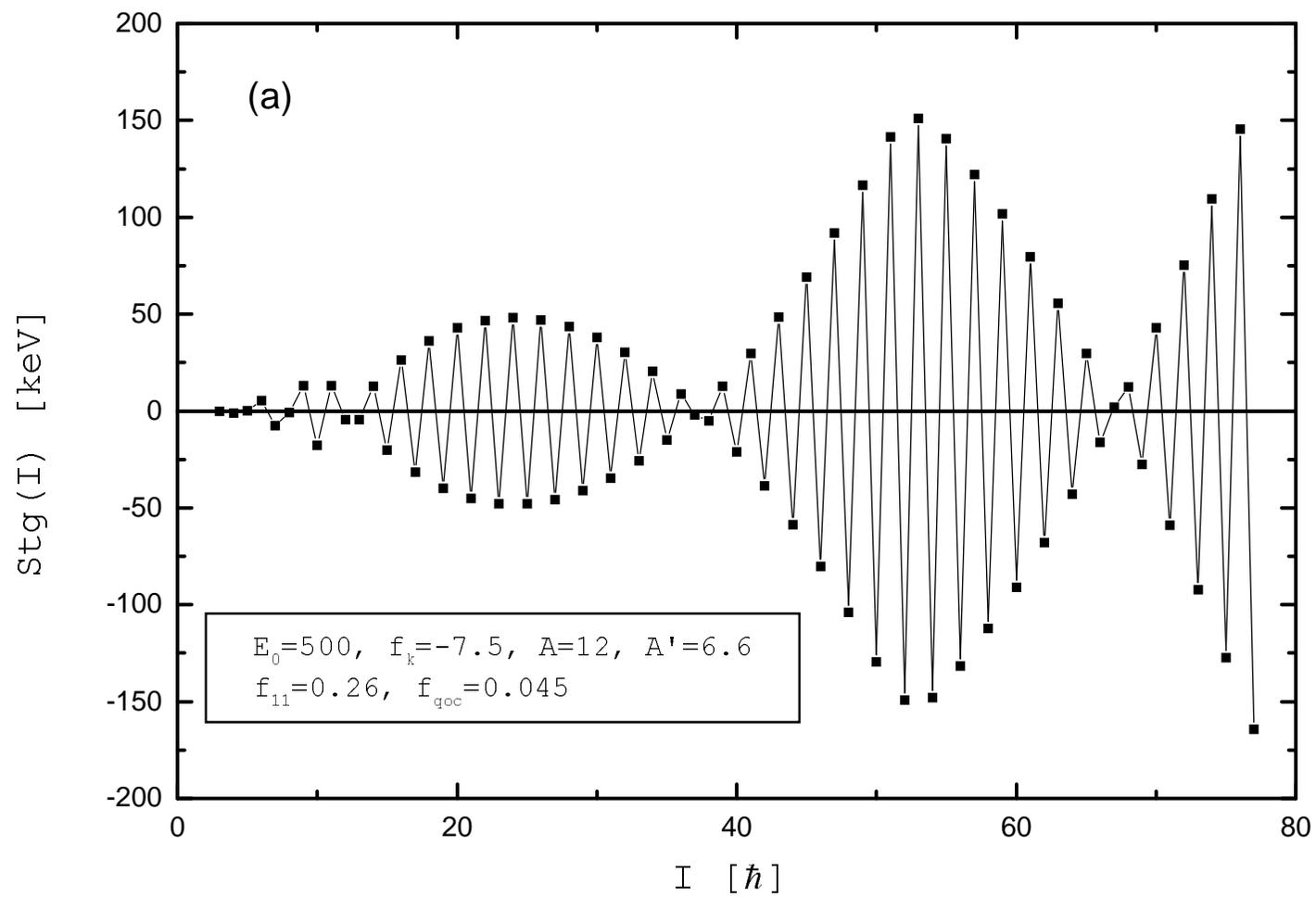
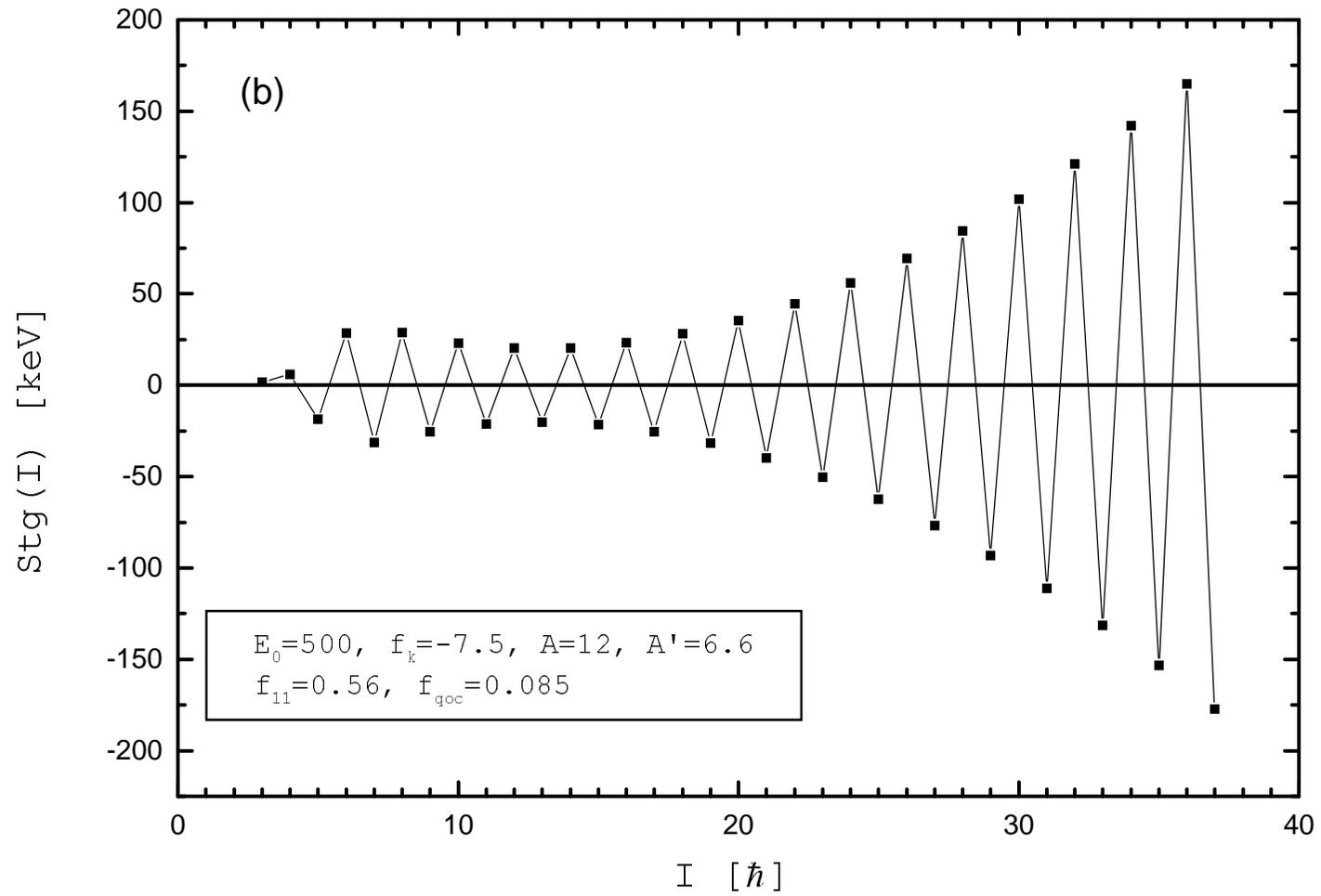
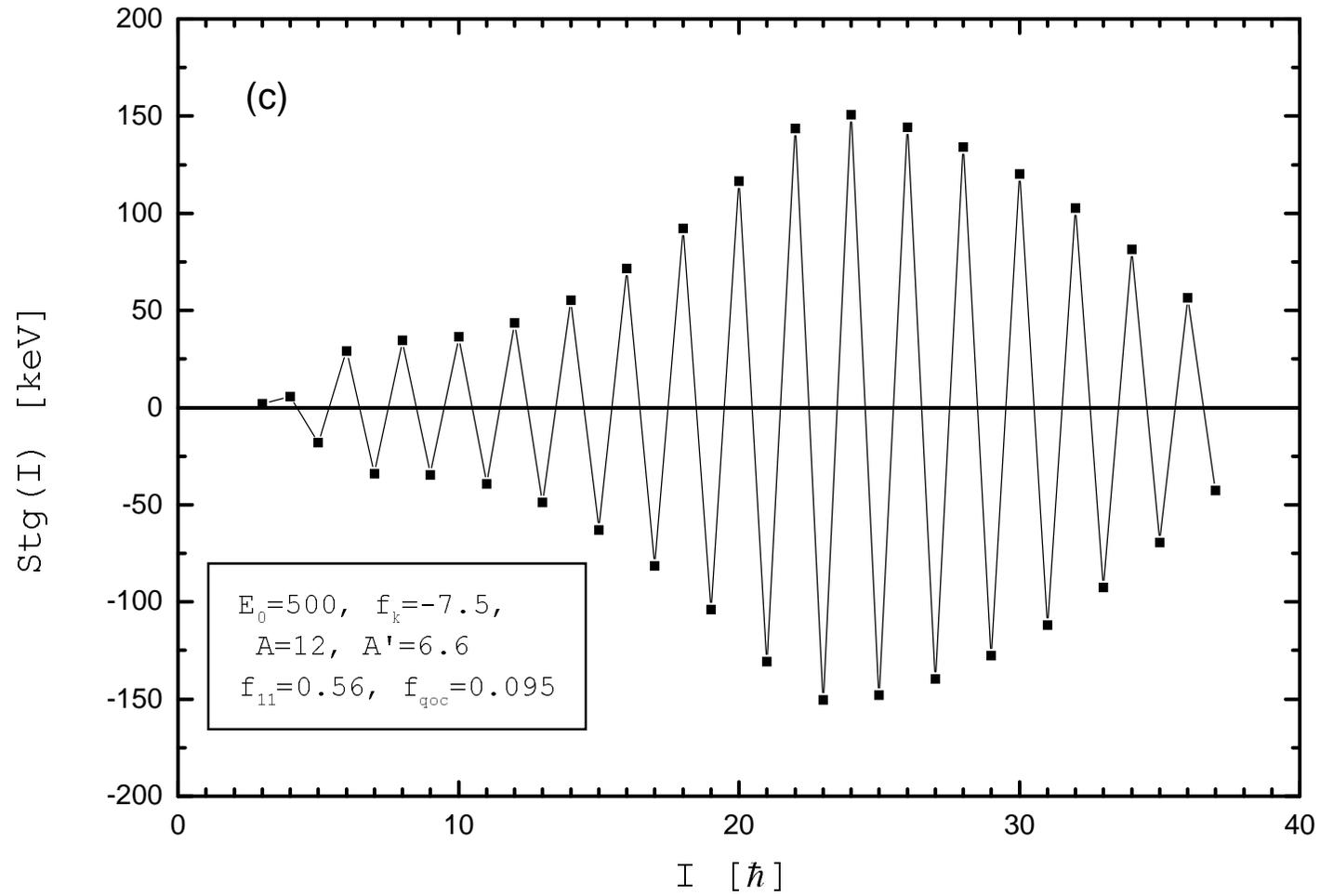
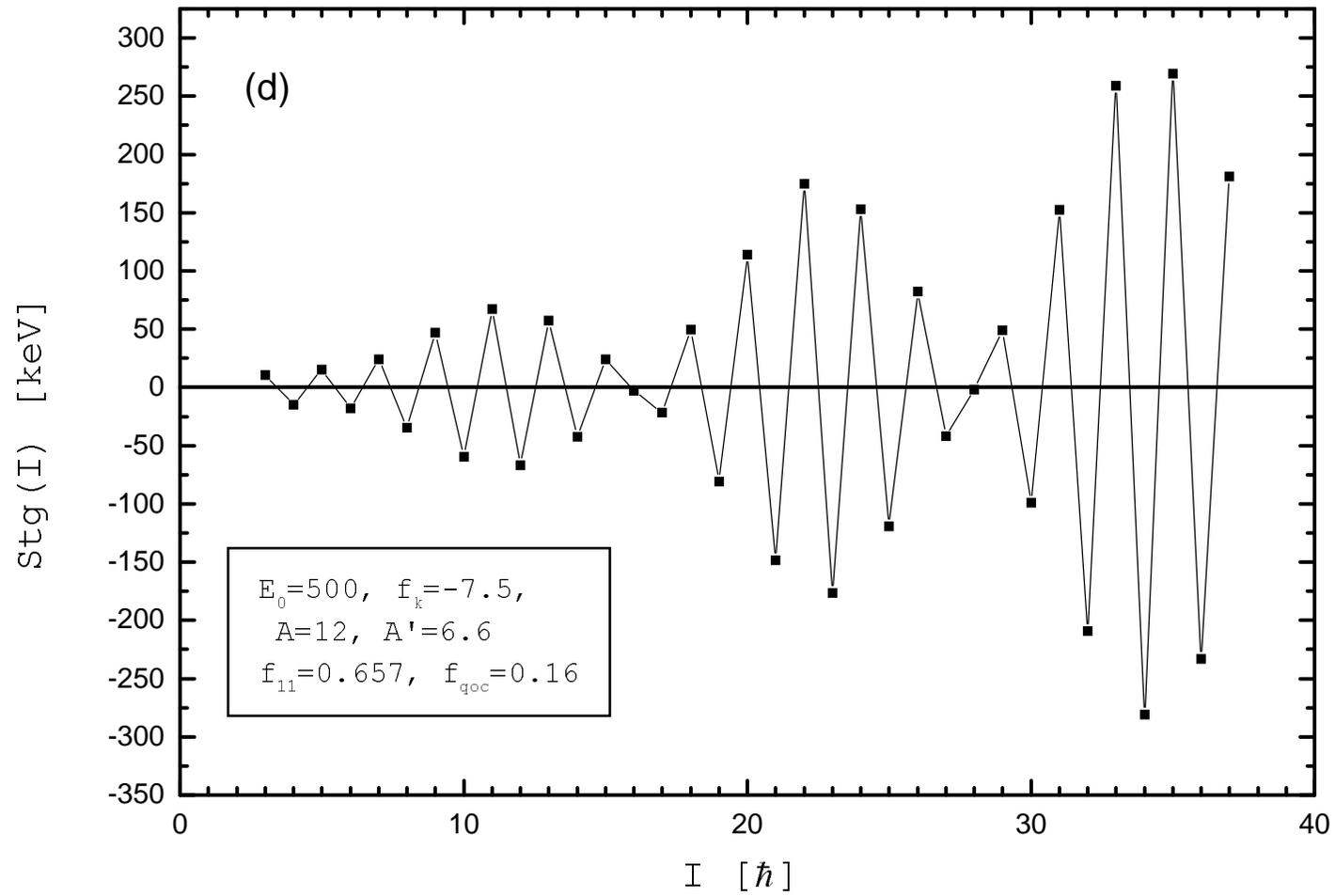
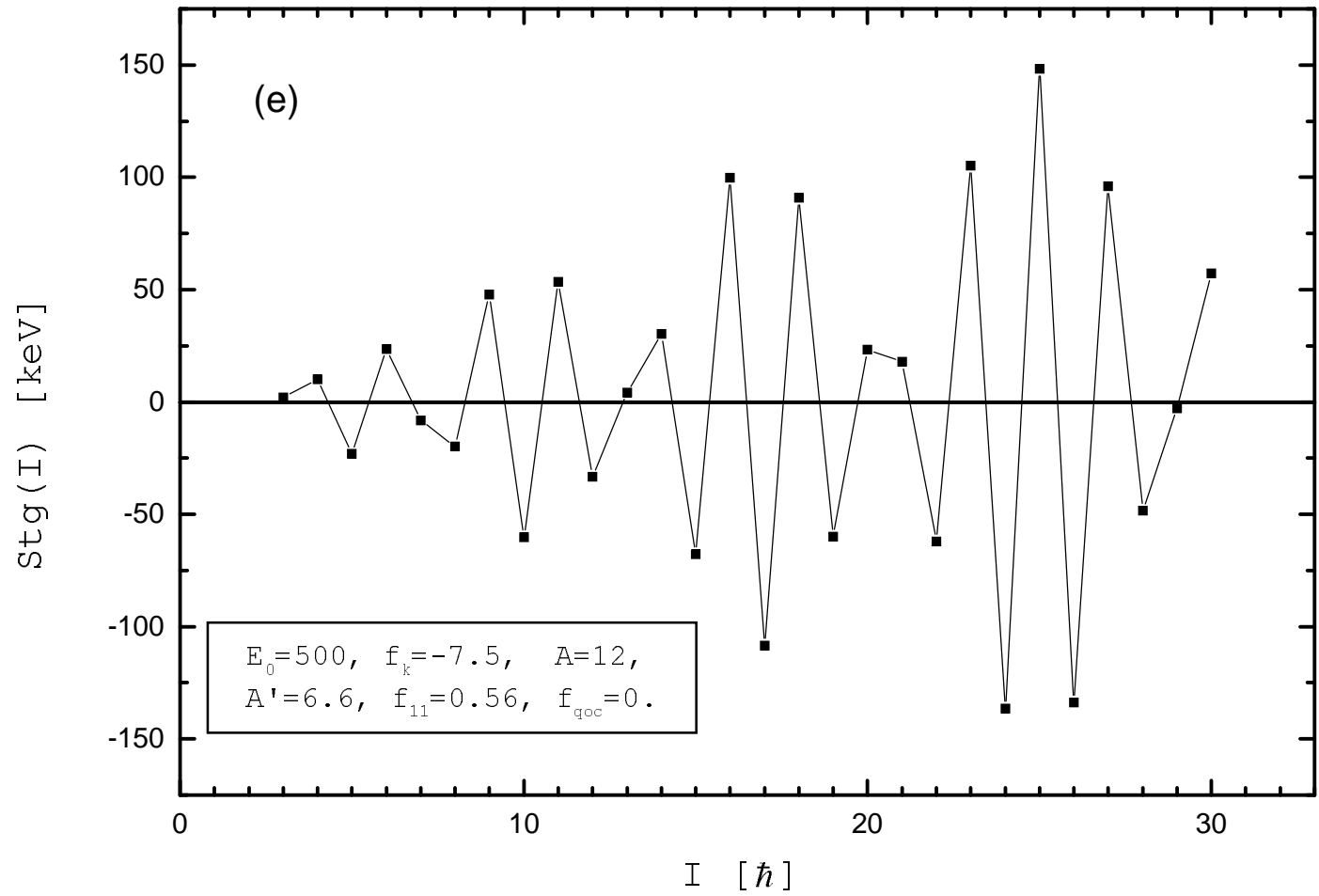
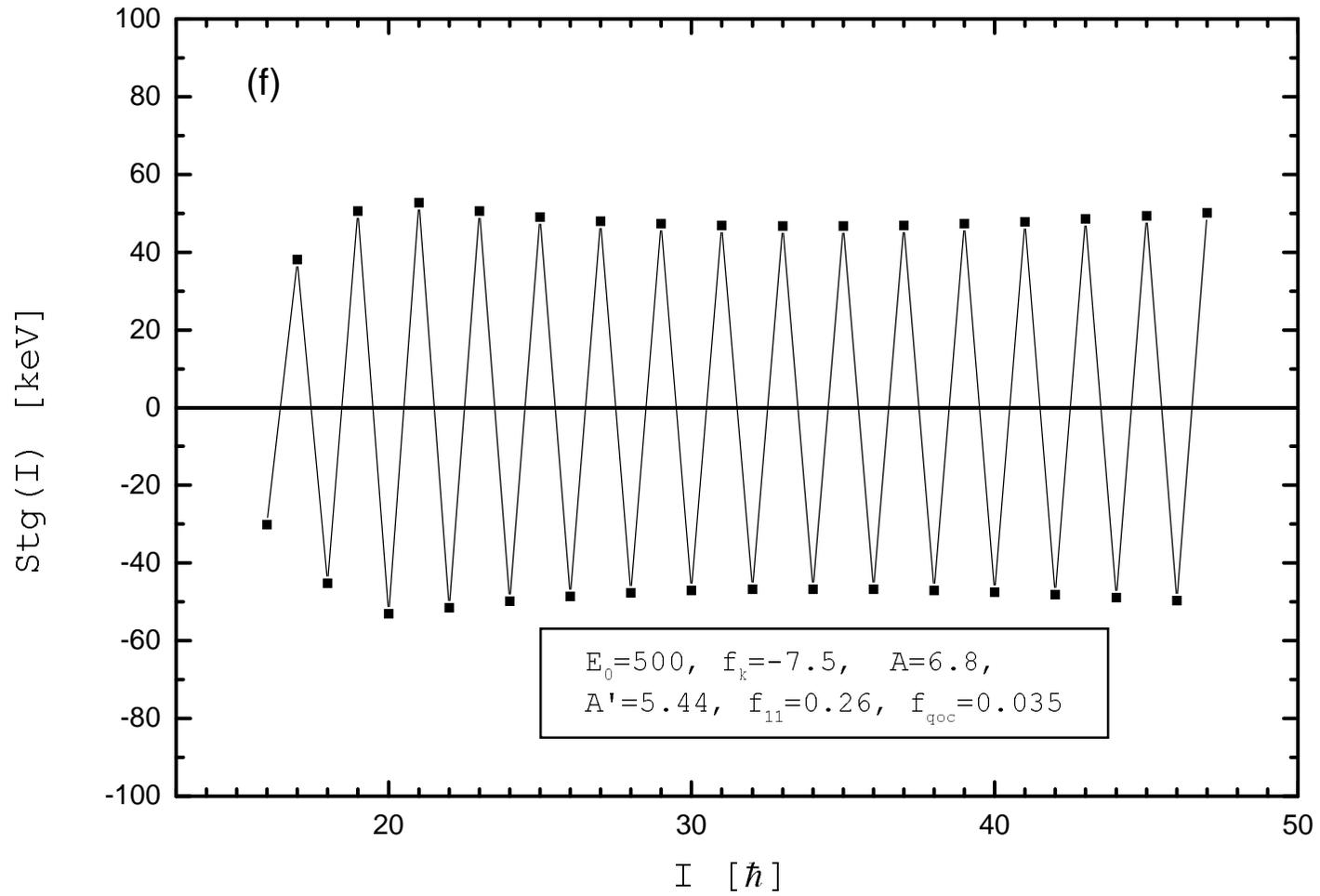
{\bf Figure 3.} Odd--even  staggering patterns (calculated using
Eq.~(\protect\ref{stag})~) as functions of $I$, obtained from the
diagonal Hamiltonian of Eq. (\ref{Hdiag}) for several different
sets of model parameters, given on the figures. Part (b)
corresponds to the same set of parameters as Table 1 and Figs 1 and 2.
The first four parameters remain the same in all parts of the figure,
except (f).
\bigskip

\noindent
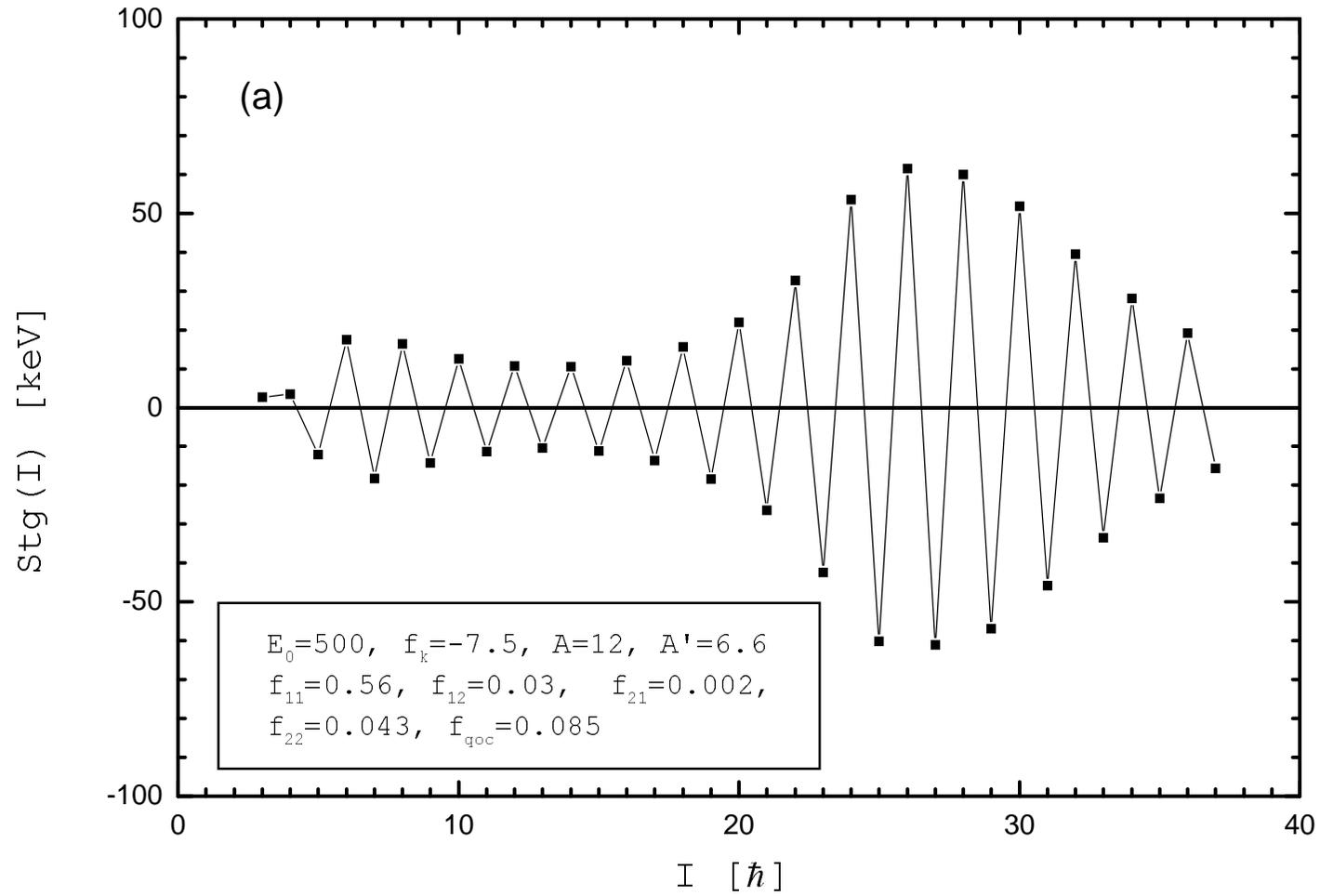
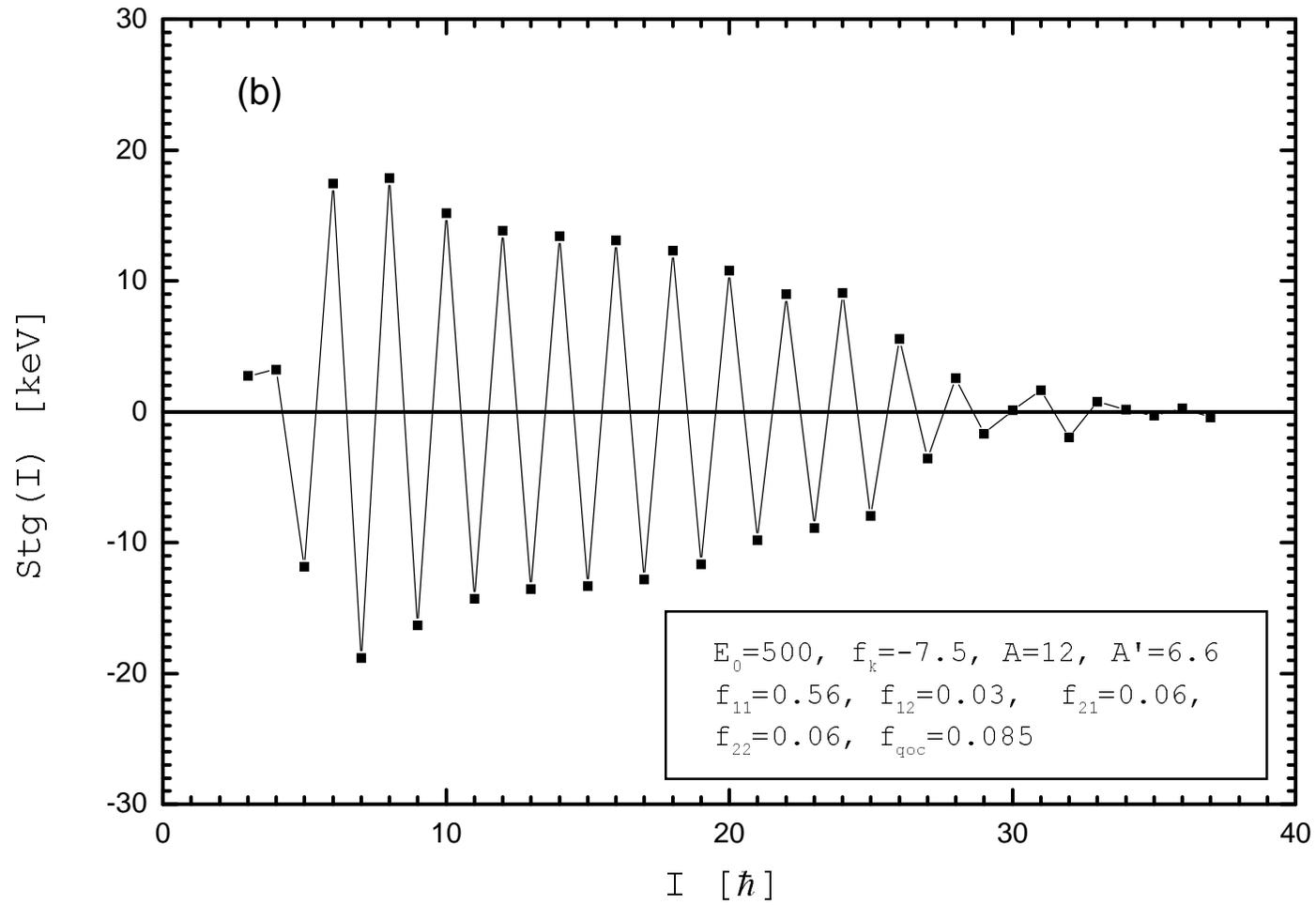
{\bf Figure 4.} Odd--even staggering patterns (calculated using
Eq.~(\protect\ref{stag})~) obtained by adding the three
non-diagonal terms $\hat{H}_{F_{1}(2)}$ (Eq.~(\ref{HF12})~),
$\hat{H}_{F_{2}(1)}$ (Eq.~(\ref{HF21})~), and $\hat{H}_{F_{2}(2)}$
(Eq.~(\ref{HF22})~) to the diagonal Hamiltonian of Eq.
(\ref{Hdiag}), for  two different sets of model parameters, given
on the figures. The first six parameters are the same as in Figs 1,
2 and 3(b).
\bigskip

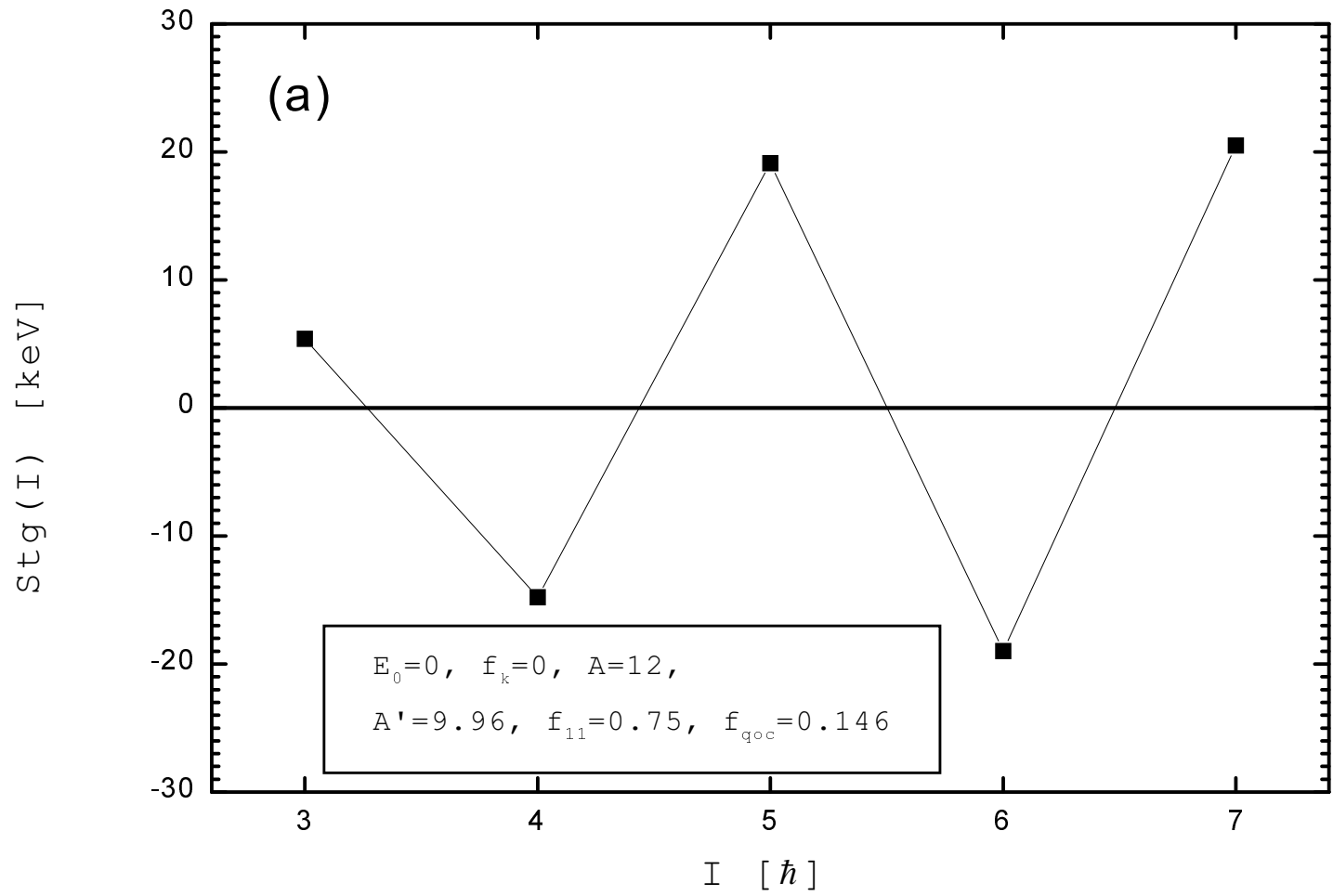
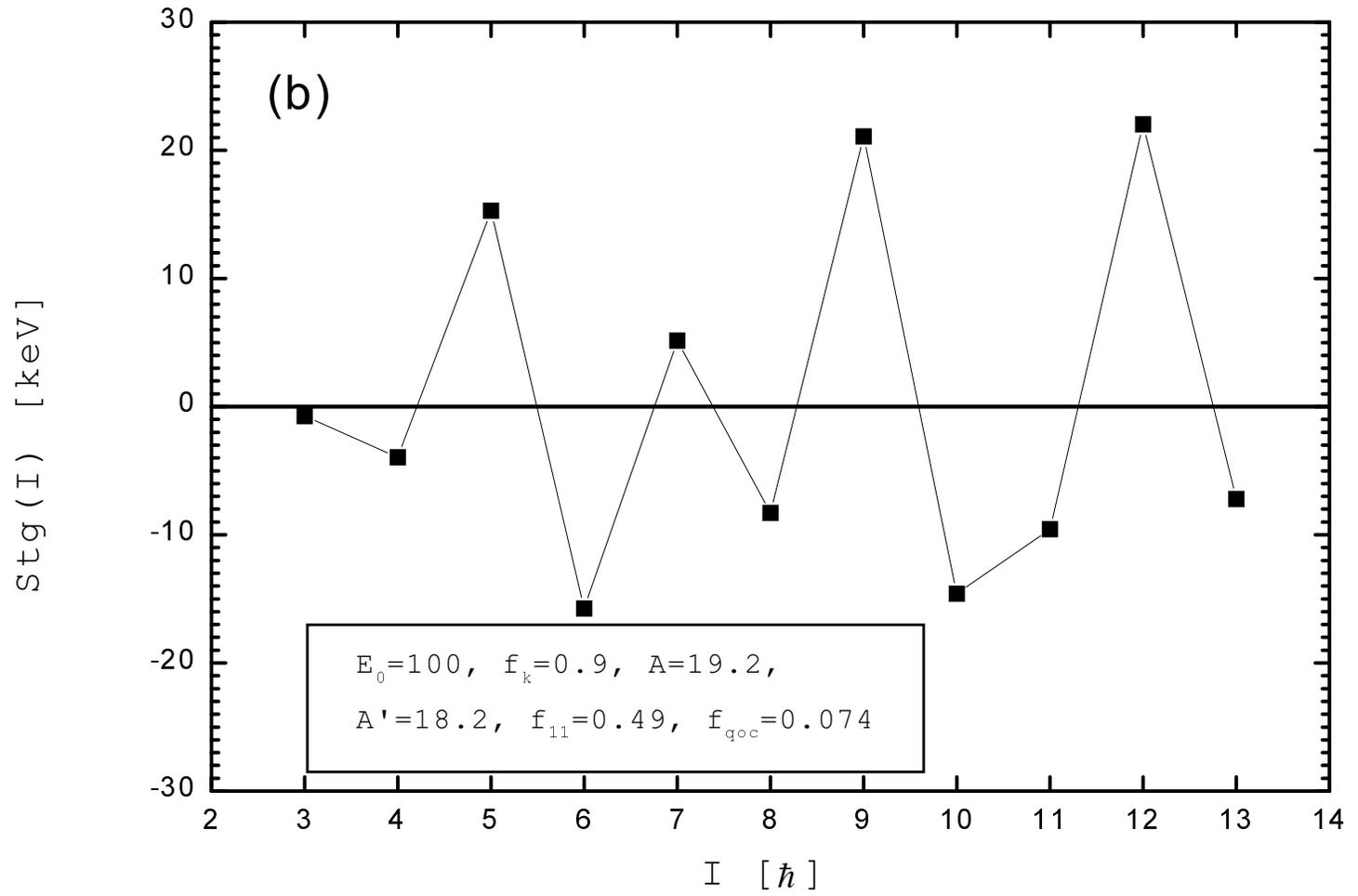
\noindent {\bf Figure 5.} Same as Fig. 3, but with the restriction
$K\leq 3$ imposed in the angular momentum regions:
(a) $I\leq 10$ and (b) $I\leq 16$.
The parameter sets correspond to Tables 2 and 3 respectively.

\end{document}